\documentclass[12pt,a4paper]{revtex4}
\usepackage{eurosym}
\usepackage[a4paper, total={6.5in, 9in}]{geometry}
\usepackage{float}
\usepackage{graphics}
\usepackage{graphicx}
\usepackage{amsmath}
\usepackage[title]{appendix}
\usepackage{subfigure}

\begin{document}

\title{On the long and short-range adhesive interactions in viscoelastic contacts}
\author{G. Violano}
\email{guido.violano@poliba.it}

\affiliation{Department of Mechanics, Mathematics and Management, Polytechnic University
of Bari, Via E. Orabona, 4, 70125, Bari, Italy}

\author{L. Afferrante}
\affiliation{Department of Mechanics, Mathematics and Management, Polytechnic University
of Bari, Via E. Orabona, 4, 70125, Bari, Italy}

\begin{abstract}
Recently, tribologists have shown increasing interest in rate-dependent phenomena occurring in viscoelastic fractures. However, in
some cases, conflicting results are obtained despite the use of similar
theoretical models. For this reason, we try to shed light on the effects
that long and short-range adhesion has on the pull-off force in the contact
of viscoelastic media by exploiting a  recently developed numerical model.

We find that, in the limit of long-range adhesion, the unloading velocity has little effect on the pull-off force, which is close to the value predicted by Bradley for rigid bodies. In such case, the detachment process is
characterized by a uniform bond-breaking of the contact area, and viscous
dissipation involves the bulk material.

For medium(short)-range adhesion, the pull-off force is instead a monotonic
increasing function of the pulling velocity and, at high speeds, reaches a plateau that is a function of the adiabatic surface energy. In this case,
the detachment process is similar to the opening of a circular crack, and
viscous dissipation is localized at the contact edge.

\end{abstract}
\maketitle

\section{Introduction}

In the 1970s, two pioneering contact mechanics theories were formulated to
study the adhesion between elastic spheres. First, Johnson, Kendall \&
Roberts (JKR) \cite{JKR1971} exploited a thermodynamic approach to include
adhesion in the Hertz theory. JKR solution assumes infinitely short-range
adhesive interactions inside the contact area, that cause a deformation of
the Hertzian contact profile. Later, Derjaguin, Muller \& Toporov (DMT) \cite%
{DMT1975} proposed a different solution where weak long-range adhesive
interactions are localized outside the contact area, without deforming the
Hertzian gap.

Both JKR and DMT theories predict a pull-off force, i.e., the maximum
tensile force reached during the detachment, that is independent of the
elastic properties of the half-space and equal to $1.5\pi \Delta \gamma R$
and $2\pi \Delta \gamma R$, respectively, being $R$ the radius of the sphere
and $\Delta \gamma $ the adiabatic surface energy.

Tabor \cite{Tabor1977} first clarified that JKR theory is accurate in the limit of high values of a dimensionless parameter $\mu $ (known as Tabor parameter), namely for soft materials with high surface energy and radii of
curvature. On the contrary, DMT\ theory is accurate in the limit of $\mu \ll
1$. Notice DMT pull-off force is the same calculated by Bradley \cite%
{Bradley1932} in the limit of rigid bodies.

Maugis \cite{Maugis1992}, exploiting the Dugdale cohesive law, formulated a
theory, known as Maugis-Dugdale (MD) theory, which allows to capture the
JKR-DMT\ transition in terms of $\mu $. Therefore, he showed that, moving
from JKR\ to DMT\ limit, the pull-off force ranges in between $1.5\pi \Delta
\gamma R$ and $2\pi \Delta \gamma R$, as also confirmed by a successive
numerical study \cite{Greenwood1997}.

In modern adhesive systems, soft materials are widely used to enhance
adhesive features. Such materials exhibit rate-dependent adhesion as a
consequence of their intrinsic viscoelasticity, while JKR, DMT, and MD
theories cannot capture rate effects and viscous dissipation. In fact,
experiments have shown that the viscoelastic pull-off force may be order of
magnitudes larger than the value predicted by \textit{elastic} adhesion
theories (see, for example, Ref. \cite{Violano2021a}).

Recently, Das \& Chasiotis (DC) \cite{DC2021} proposed to extend MD model to the
case of a rigid sphere indenting a \textit{viscoelastic} half-space. In their
works, unloading starts from a relaxed state of the viscoelastic material
and is performed at fixed rate of the applied force.

In the limit of short-range adhesive interactions, DC model predicts the
pull-off force to be a monotonic increasing function of the unloading rate.
If $F_{\mathrm{PO,0}}$ denotes the elastic pull-off force, the maximum
pull-off force that can be reached at high unloading rate is $F_{\mathrm{%
PO,\max }}=(F_{\mathrm{PO,0}}E_{\infty })/E_{0}$, being $E_{\infty }$ and $%
E_{0}$ the instantaneous and relaxed moduli of the viscoelastic material,
respectively. However, such value cannot be reached for any value of the
Tabor parameter (see, for example, Refs. \cite{ViolanoSIZE, Muser2021, Ciavarella2021}).

Moreover, when long-range adhesive interactions are predominant, namely when $\mu \rightarrow 0$,
DC model continues to predict $F_{\mathrm{PO,\max }}=(F_{\mathrm{PO,0}%
}E_{\infty })/E_{0}$ at high unloading rate, while in Ref. \cite{Ciavarella2021} it is observed that $F_{\mathrm{PO,\max }}\rightarrow F_{\mathrm{PO,0}}$ independently of the value of the unloading rate.

In this work, we try to clarify such qualitative and quantitative
discordance by performing contact simulations with the fully deterministic
model developed in Ref. \cite{Afferrante2022}, where adhesive interactions
are modeled with Lennard-Jones based potential and viscoelasticity with the
standard linear solid.

\section{Statement of the problem}

Consider a rigid sphere, with radius of curvature $R$, pulled off from a
viscoelastic substrate. Detachment starts from static equilibrium conditions
as the sphere initially approaches the substrate at low speed up to a
maximum penetration $\delta _{\max }$. Retraction is then performed at a
constant speed $V=-d\delta /dt$.

The contact problem is solved using the finite element (FE) model presented
in Ref. \cite{Afferrante2022}, at which the reader is referred for further
details. We here recall that: i) interface interactions are modelled by a
traction-gap law based on Lennard-Jones potential, according to the \textit{%
proximity approximation} (i.e., the force acting between two bodies is
expressed in terms of the force between two semi-infinite planes); ii) the
viscoelastic behaviour of the substrate is modelled with the standard linear
solid according to the Maxwell representation.

We are interested in studying how the pull-off force is affected by the
retraction speed $V$, in the cases of short, medium, and long range adhesive
interactions. In the elastic case, the Tabor parameter is defined as%
\begin{equation}
\mu =\frac{R^{1/3}}{\epsilon }\left[ \frac{\Delta \gamma (1-\nu ^{2})}{E_{0}}%
\right] ^{2/3}\text{,}  \label{Tabor}
\end{equation}%
where $E_{0}$ is Young's modulus of the substrate, $\epsilon $ is the range
of action of attractive forces, and $\nu $ is Poisson's ratio.

For a viscoelastic material, Young's modulus $E(\omega )$ is a function of
the frequency of excitation and hence varies during the detachment process.
Viscoelastic modulus $E(\omega )\rightarrow E_{0}$ only when detachment
occurs in quasi-static conditions, namely when $V\rightarrow 0$. As a
result, while in the elastic case we are sure that $\mu $ uniquely describes
the trend of the pull-off force, in the viscoelastic case this it is not
necessarily true. However, we can reasonably assume that long-range adhesion
is expected for low values of the adiabatic surface energy $\Delta \gamma $,
resulting in $\mu \ll 1$, while medium and long-range adhesion occurs at
higher $\Delta \gamma $. In the following, we shall discuss our results in
terms of $\Delta \gamma $ rather than $\mu $.

\section{Results and Discussion}

All plots are given for $E_{\infty }/E_{0}=10$ and $\nu \approx 0.5$, and
in terms of the following dimensionless quantities: $\hat{F}_{\mathrm{PO}%
}=F_{\mathrm{PO}}/(\pi \Delta \gamma R)$; $\Delta \hat{\gamma}=\Delta \gamma
/(E_{0}R)$; $\hat{V}=V\tau /\epsilon $, being $\tau $\textbf{\ }the
relaxation time of the viscoelastic material. Furthermore, the unloading
phase started after reaching the penetration $\hat{\delta}_{\max }=\delta
/\epsilon =17.7$.

\begin{figure}[tbp]
\begin{center}
\includegraphics[width=12.0cm]{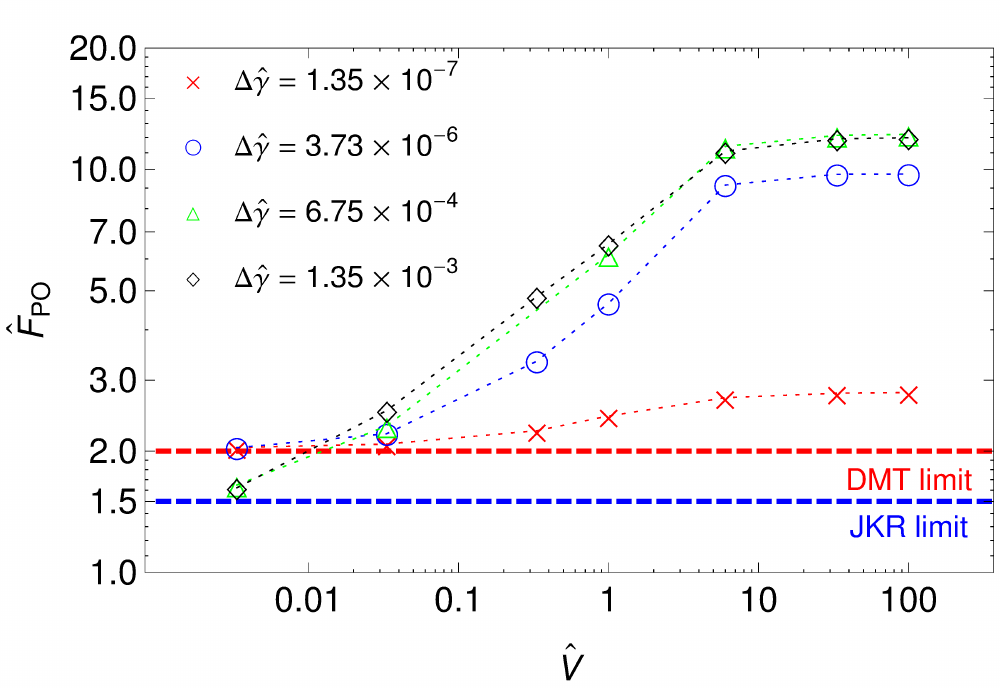}
\end{center}
\caption{The normalized pull-off force $\hat{F}_{\mathrm{PO}}$ in terms of
the dimensionaless pulling velocity $\hat{V}$. Finite element calculations
(markers) are shown for $\Delta \hat{\protect\gamma}=1.35\times 10^{-7}$, $%
3.73\times 10^{-6}$, $6.75\times 10^{-4}$, $1.35\times 10^{-3}$. DMT\ and JKR elastic limits are also shown as a reference.}
\label{FIGURE1}
\end{figure}

Figure \ref{FIGURE1} shows the normalized pull-off force $\hat{F}_{\mathrm{PO%
}}$ in terms of the pulling velocity $\hat{V}$, which is constant during the
retraction of the sphere. Results are obtained for different values of the
adiabatic surface energy $\Delta \hat{\gamma}$ and for $\hat{V}$ ranging
from $0.0033$ to $100$. First, let us consider what happens when $V\sim 0$;
for the lower values of $\Delta \hat{\gamma}$, the DMT\ limit is returned,
being $\hat{F}_{\mathrm{PO}}\sim 2$. As a result, adhesion is basically
described by long-range interactions. On the contrary, for the higher values
of $\Delta \hat{\gamma}$, $\hat{F}_{\mathrm{PO}}\sim 1.5$ and the JKR limit
is recovered being dominant the short-range adhesion. In all cases, $\hat{F}%
_{\mathrm{PO}}$ is a monotonic increasing function of $\hat{V}$ and reaches
a maximum asymptotic value $\hat{F}_{\mathrm{PO,\max }}$ at high pulling
velocities. However, notice $\hat{F}_{\mathrm{PO,\max }}$ is strongly
affected by the value of $\Delta \hat{\gamma}$.

In this regard, fracture mechanics and cohesive zone models (see, for
example, Refs. \cite{greenwood2004, PB2005}) suggest that $F_{\mathrm{%
PO,\max }}/F_{\mathrm{PO,0}}=E_{\infty }/E_{0}$, being $F_{\mathrm{PO,0}}$
the pull-off force in the elastic limit. Our calculations, instead, returns $%
F_{\mathrm{PO,\max }}/F_{\mathrm{PO,0}}\sim 0.8E_{\infty }/E_{0}$ for $%
\Delta \hat{\gamma}=$ $1.35\times 10^{-3}$ and $F_{\mathrm{PO,\max }}/F_{%
\mathrm{PO,0}}\sim 0.14E_{\infty }/E_{0}$ for $\Delta \hat{\gamma}=$ $%
1.35\times 10^{-7}$. These results agree with the theoretical findings given
in Ref. \cite{Ciavarella2021}, where it is suggested that there is an upper
bound of the pull-off force that depends on the Tabor parameter. However,
this is not the only reason for which  $F_{\mathrm{PO,\max }}/F_{\mathrm{PO,0%
}}$ is less than $E_{\infty }/E_{0}$; indeed, in a recent study \cite%
{ViolanoSIZE}, we have shown that geometric and finite-size effects
influence $F_{\mathrm{PO,\max }}$, which decreases when retraction starts
from smaller initial contact radii $a_{\max }$.

\begin{figure}[tbp]
\begin{center}
\includegraphics[width=12.0cm]{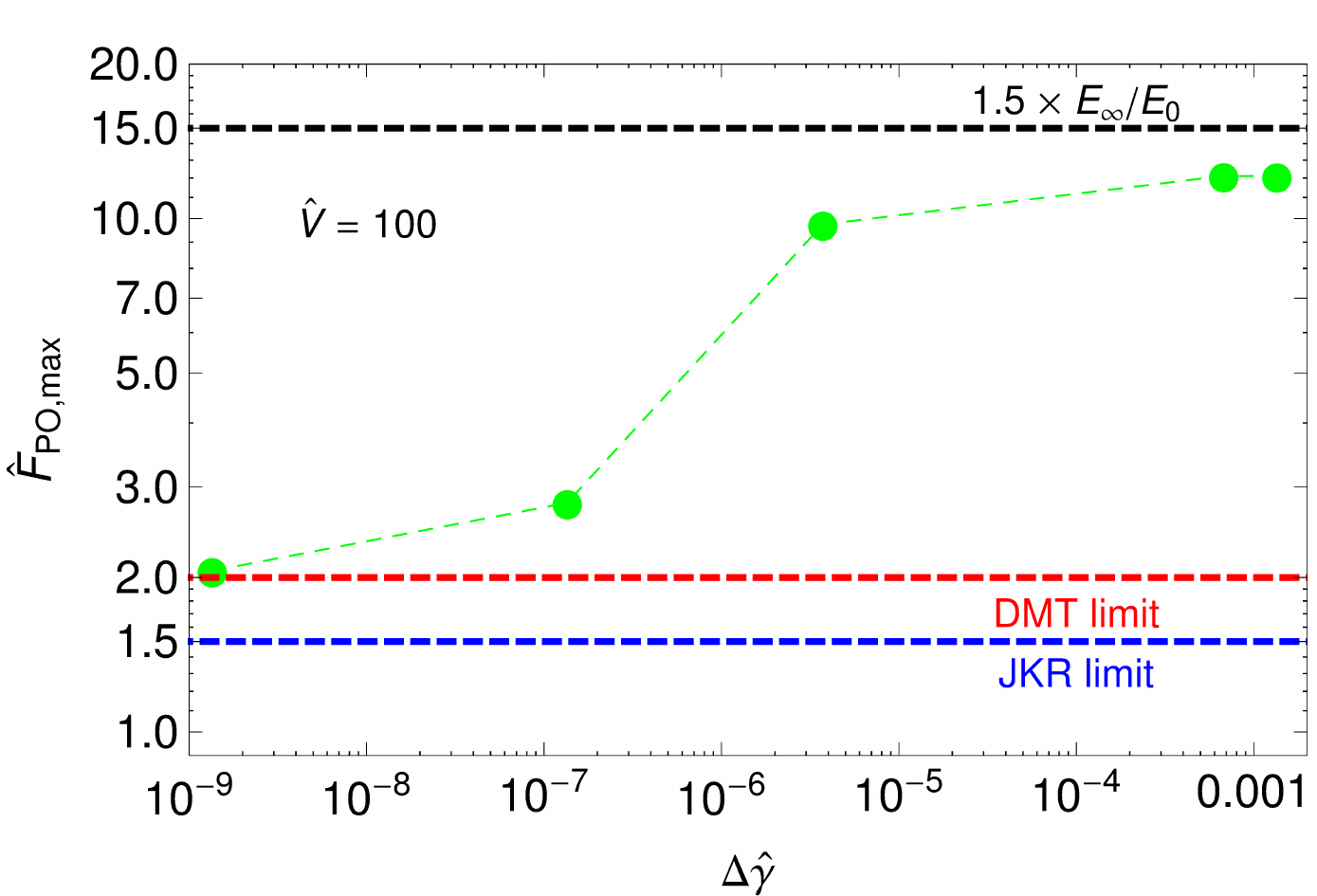}
\end{center}
\caption{The maximum pull-off force $\hat{F}_{\mathrm{PO,\max }}$ in terms of the dimensionaless adiabatic surface energy $\Delta \hat{\protect\gamma}$. $\hat{F}_{\mathrm{PO,\max }}$ is calculated at
high pulling speed ($\hat{V}=100$). Green markers refer to finite element
calculations, red and blue dashed lines are DMT\ and JKR elastic
limits, respectively, and black dashed line is the theoretical limit
predicted at high speeds by fracture mechanics models.}
\label{FIGURE2}
\end{figure}

To highlight the effect of the adiabatic surface energy on the maximum pull-off force, Fig. \ref{FIGURE2}\textbf{\ }shows $\hat{F}_{\mathrm{PO,\max}}$ in terms of $\Delta \hat{\gamma}$ and for $\hat{V}=100$. Results disagree with DC findings \cite{DC2021}, where $\hat{F}_{\mathrm{PO,\max }}$ is predicted to increase when we move from short-range to long-range adhesive interactions for all rates of unloading (see Figure 6C in Ref. \cite{DC2021}). Therefore, in DC calculations, $\hat{F}_{\mathrm{PO,\max }}$ is surprisingly found to be a decreasing function of the Tabor parameter.

However, we have to observe that our outcomes agree with results given in Ref. \cite{Ciavarella2021}, where a similar model to DC one, based on Maugis-Dugdale approach, is adopted. Our results are also in agreement with recent numerical findings of Muser \& Persson (MP) \cite{Muser2021}, who performed calculations on the detachment of a rigid cylinder from a viscoelastic substrate. They found a decrease in $\hat{F}_{\mathrm{PO,\max }}$ when reducing the Tabor parameter $\mu $. However, they changed the size of the radius $R$ of the indenter maintaining the same value of $\Delta \hat{\gamma}$. In such conditions, size effects due to $a_{\max }$ may arise, leading to additional variations in the pull-off force \cite{ViolanoSIZE}; moreover, at small scales, a transition of the detachment mode from crack propagation to uniform bond breaking can also occur \cite{ViolanoSIZE,Muser2021,Gao2004}.

In addition, Van Dokkum et al. \cite{VanDokkum2021} observed, in a similar problem, the normalized effective surface energy $\Delta \gamma _{\mathrm{eff}}/\Delta \gamma $ (or equivalently $\hat{F}_{\mathrm{PO}}$) decreases with the Tabor parameter. Furthermore, Jiang et al. \cite{Jiang2021}, performing FE calculations on the pull-off force of a spherical indenter from a soft viscoelastic stamp, observed a reduction of $\hat{F}_{\mathrm{PO}}$ when decreasing $\Delta \hat{\gamma}$. Similarly to our approach, they changed the value of $\Delta \hat{\gamma}$ maintaining the same maximum penetration $\delta _{\max }$.

Our results find confirmation also in the experimental measurements of Ahn \& Shull (AS) \cite{Ahn1998}, who performed axisymmetric adhesion tests between a hemispherical elastomeric cap and a variety of flat substrates with different values of interface energy. Moving from the JKR\ formalism, they used a standard fracture mechanics approach to estimate the effective surface energy $\Delta \gamma _{\mathrm{eff}}$ according to the phenomenological equation $\Delta \gamma _{\mathrm{eff}}=\Delta \gamma \lbrack 1+f(V_{\mathrm{c}})]$ \cite{MB1980}, where the quantity $f(V_{\mathrm{c}})$ takes into account the increase in surface energy due to viscous dissipation and depends only on the contact line velocity $V_{\mathrm{c}}=-da/dt$ (provided the process is assumed occurring at constant temperature).

\begin{figure}[tbp]
\begin{center}
\includegraphics[width=18.0cm,trim={3cm 1cm 3cm 0},clip]{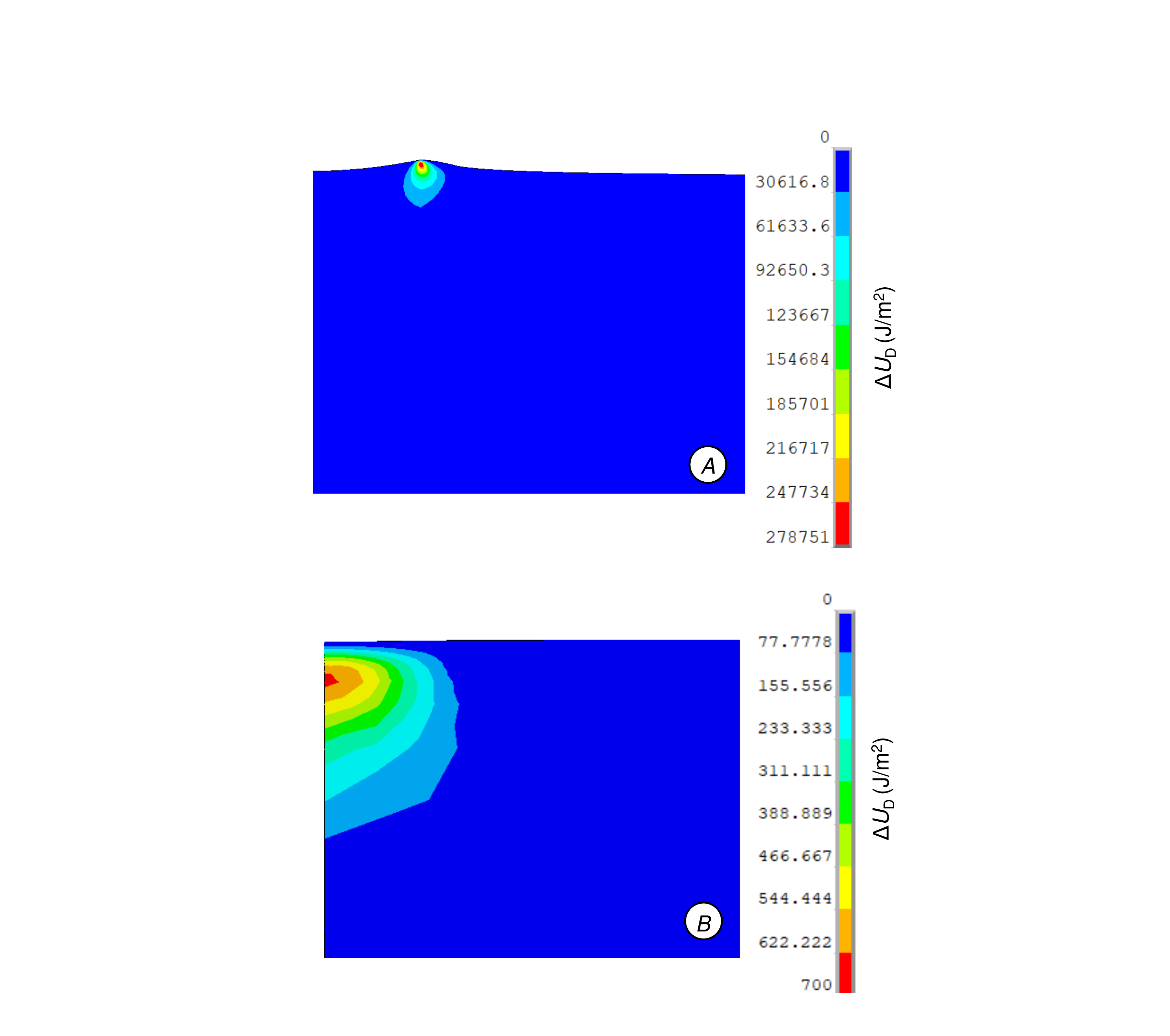}
\end{center}
\caption{The increase in the dissipated energy density at
pull-off. Results are given for a pulling speed $\hat{V}=1$, with $\Delta 
\hat{\protect\gamma}=$ $1.35\times 10^{-3}$ (A) and $\Delta \hat{\protect%
\gamma}=1.35\times 10^{-7}$ (B).}
\label{FIGURE3}
\end{figure}

\begin{figure}[tbp]
\begin{center}
\includegraphics[width=12.0cm]{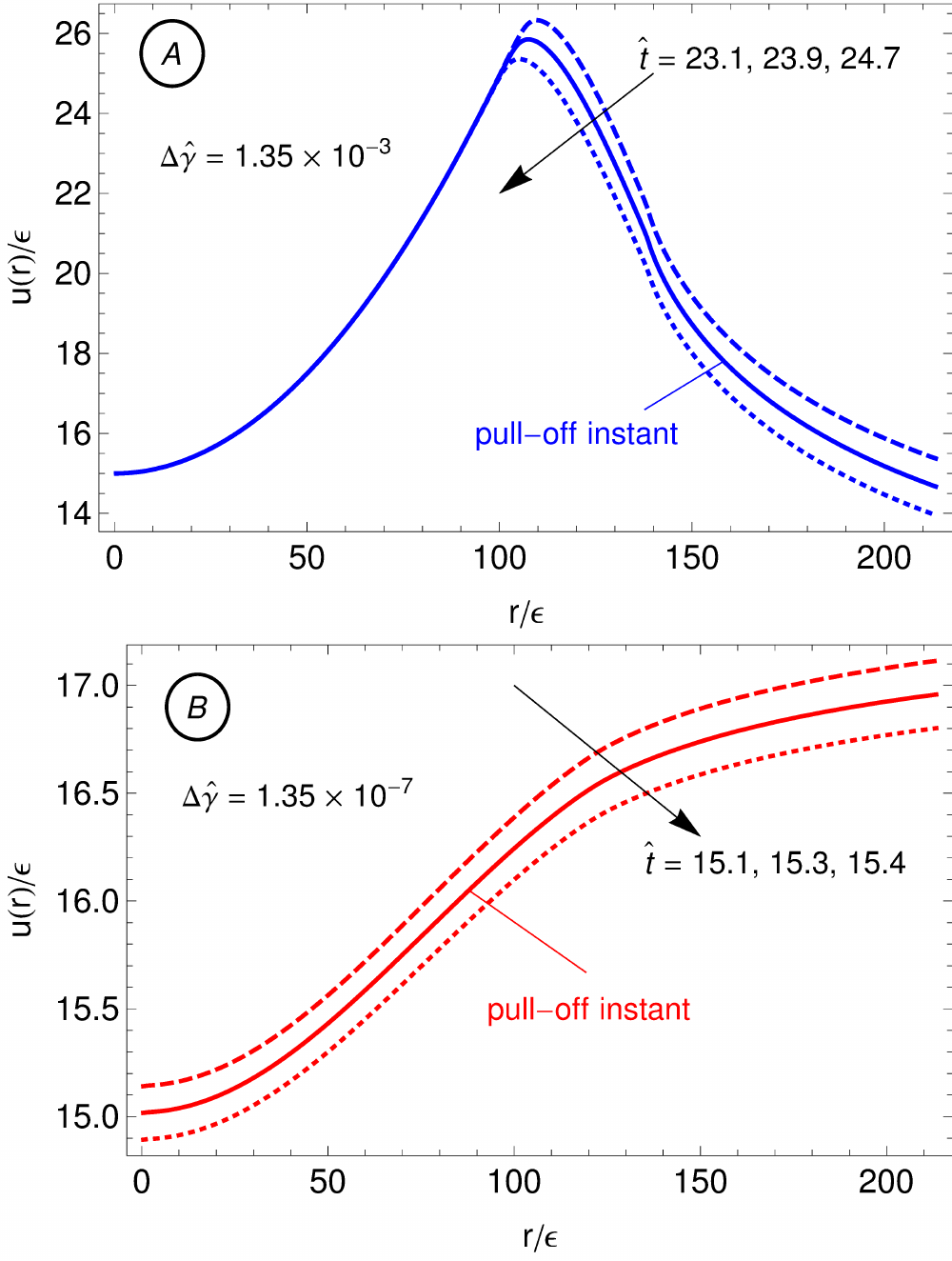}
\end{center}
\caption{Displacement fields in the vicinity of the pull-off instant, for a pulling speed $\hat{V}=1$, with $\Delta \hat{\protect\gamma}=$ $1.35\times 10^{-3}$ (A) and $\Delta \hat{\protect\gamma}=1.35\times 10^{-7}$ (B). The dimensionless time $\hat{t}=t/\protect\tau$ increases in the direction of the arrow. The displacements are given net of the rigid one.}
\label{FIGURE4}
\end{figure}

Figures \ref{FIGURE3}A-B show the increase in the dissipated energy density $%
{\Delta }U_{\mathrm{D}}$ at pull-off, for $\hat{V}=1$ and $\Delta \hat{\gamma%
}$ equal to $1.35\times 10^{-3}$ and $1.35\times 10^{-7}$, respectively. For
high values of the adiabatic surface energy (Fig. \ref{FIGURE3}A),
viscous dissipation is localized at the edge of contact, suggesting that the detachment process is similar to the mechanism of opening of a circular crack \cite{Afferrante2022}. For low $\Delta \hat{\gamma}$ (Fig. \ref{FIGURE3}B), a much lower viscous dissipation is observed. It mainly involves the bulk material, suggesting the occurrence of a different debonding mechanism, as clarified in Figs. \ref{FIGURE4}A-B, where the normal displacement (net of the rigid one) is shown at three different times close to the pull-off instant (namely the instant at which the tensile force is maximum).

Results are given for the two cases corresponding to Figs. \ref{FIGURE3}A-B. For $\Delta \hat{\gamma}=$ $1.35\times 10^{-3}$, the debonding mechanism is clearly related to crack propagation; for $\Delta \hat{\gamma}=$ $1.35\times 10^{-7}$, the displacement field instead moves homogeneously leading to a quasi-uniform bond breaking.

\section{Conclusions}

In this paper, we investigate the effect of the adhesive
interactions on the pull-off force of viscoelastic bodies when the
detachment process starts from a relaxed equilibrium state of the material.

In the limit of long-range adhesion ($\Delta \gamma \rightarrow
0$), quasi-uniform bond-breaking describes the mechanism of detachment, and viscous dissipation occurs in the bulk material. As a result, the pull-off force is not affected by the pulling speed and tends to the rigid limit of Bradley \cite{Bradley1932}.

When medium(short)-range interactions occur, the detachment process is similar to that of a propagating circular crack. In this case, the pull-off force is a monotonic increasing function of the pulling speed. At high speeds, the pull-off load reaches a plateau, which depends on $\Delta \gamma$.

Therefore, we can state that the pull-off force is limited by the
strength of adhesion ($\Delta \gamma $), rate of unloading and finite-size effects ($a_{\max }$). This is the reason why the ratio between the actual pull-off force and its elastic value does not approach the theoretical limit $E_{\infty }/E_{0}$ at high pulling speeds.

Such outcomes also have important implications in problems of viscoelastic fracture mechanics.

\section*{ACKNOWLEDGEMENTS}
G.V. and L.A. acknowledge support from the Italian Ministry of Education, University and Research (MIUR) under the programme
"Departments of Excellence" (L.232/2016).

\section*{DECLARATIONS}

\subsection*{Conflict of interest}
The authors declare that there are no conflicts of
interests within the outlined work.

\subsection*{Fundings}
The authors declare that no funds, grants, or other support were received during the preparation of this manuscript.

\end{document}